\documentclass[prl,superscriptaddress,twocolumn,showpacs]{revtex4}
\usepackage{graphicx,amssymb,amsmath,bbm,amsfonts,amsbsy}
\newcommand{\sab}{{\scriptstyle AB}}
\newcommand{\smab}{{\scriptstyle AB}}
\newcommand{\smb}{{\scriptstyle B}}
\newcommand{\sma}{{\scriptstyle A}}

\newcommand{\scab}{{\scriptstyle A|B}}

\newcommand{\smx}{{\scriptstyle X}}
\newcommand{\smt}{{\scriptstyle T}}
\newcommand{\smm}{{\scriptstyle M}}
\newcommand{\smbx}{{\scriptstyle \boldsymbol{X}}}
\newcommand{\smbz}{{\scriptstyle \boldsymbol{Z}}}

\newcommand{\X}{\boldsymbol{X}}
\newcommand{\Y}{\boldsymbol{Y}}
\newcommand{\Z}{\boldsymbol{Z}}
\newcommand{\id}{\mathbbm{I}}

\newcommand{\be}{\begin{equation}}
\newcommand{\ee}{\end{equation}}
\newcommand{\bae}{\begin{eqnarray}}
\newcommand{\eae}{\end{eqnarray}}

\begin{document}
%%%%%%%%%%%%%%%%%%%%%%%%%%%%%%%%%%%%%%%%%%%%%%%%%%%%%%%%%%%%%
\title{Gaussian quantum discord}
\author{Paolo Giorda}\email{giorda@isi.it}
\affiliation{ISI Foundation, I-10133 Torino, Italy}
\author{Matteo G A Paris}\email{matteo.paris@fisica.unimi.it}
\affiliation{Dipartimento di Fisica dell'Universit\`a degli Studi di
Milano, I-20133 Milano, Italy}
\date{\today}
%%%%%%%%%%%%%%%%%%%%%%%%%%%%%%%%%%%%%%%%%%%%%%%%%%%%%%%%%%%%%
\begin{abstract}
We extend the quantum discord to continuous variable systems and evaluate Gaussian quantum discord $C(\varrho)$ for bipartite Gaussian states.
In particular, for squeezed thermal states (STS),
we explicitly maximize the extractable information
over Gaussian measurements: $C(\varrho)$ is minimized
by a generalized measurement rather than a projective one.
Almost all STS have nonzero Gaussian discord:
they may be either separable or entangled if the discord is below
the threshold $C(\varrho)=1$, whereas they are all entangled above
the threshold. We elucidate the general role of state parameters in determining
the discord and discuss its evolution in noisy channels.
\end{abstract}
%%%%%%%%%%%%%%%%%%%%%%%%%%%%%%%%%%%%%%%%%%%%%%%%%%%%%%%%%%%%%
\maketitle
Quantum correlations have been the subject of intensive studies in the
last two decades, mainly due to the general belief that they are a
fundamental resource for quantum information processing tasks.  The
first rigorous attempt to address the classification of quantum
correlation from has been put forward by Werner \cite{Wer89}, who put on
firm basis the elusive concept of quantum entanglement.  A state of a
bipartite system is called entangled if it cannot be written as follows:
$\varrho_\sab = \sum p_k \varrho_{\sma k}\otimes\varrho_{\smb k}$, where
$\varrho_{\sma k}$ and $\varrho_{\smb k}$ are generic density matrices
describing the states of the two subsystems. The definition above has an
immediate operational interpretation: separable states can be prepared
by local operations and classical communication between the two parties,
whereas entangled states cannot. One might have thought that such
classical information exchange could not bring any quantum character to
the correlations in the state. In this sense separability has often been
regarded as a synonymous of classicality.  However, it has been shown
that this is not the case \cite{OZ01,HV01}.
A measure of correlations --quantum discord--
has been defined as the mismatch
between two quantum analogues of classically equivalent expression of
the mutual information. For pure entangled states quantum
discord coincides with the entropy of entanglement.  However, quantum
discord can be different from zero also for some (mixed) separable
state. In other words, classical communication can give rise to quantum
correlations. This can be understood by considering that the states
$\varrho_{\sma k}$ and $\varrho_{\smb k}$ above may be physically non
distinguishable, i.e. non-orthogonal and thus not all the information
about them can then be locally retrieved. This phenomenon has no
classical counterpart, thus accounting for the quantumness of the
correlations in separable state with positive discord.
Quantum discord has been shown to be a property hold by almost
all quantum states \cite{Ale09} and recently attracted considerable
attention \cite{Dat08,RR08,Pia08,Lan08,Dil08,Wer09}. In particular,
the vanishing of quantum discord between two systems has
been shown to be a requirement for the complete positivity of
the reduced subsystem dynamics \cite{Lid09}.
\par
While the discord is a fundamental notion allowing for the description
of the quantumness of the correlations present in the state of a quantum
system, its evaluation requires an optimization procedure over the set
of all measurements on a given subsystem, and thus attacking the general
case is a formidable task. For this reason, the original definition of
the quantum discord \cite{OZ01} involved orthogonal measurements, and
its evaluation and the study of its properties has mainly been
restricted to final dimensional systems \cite{Luo08}.  The purpose of
this Letter is to extend the notion of discord to the domain of
continuous variable systems. In the following, we focus our analysis on
bipartite systems that are described by two-mode Gaussian states and we
explore the concept of discord within the domain of generalized Gaussian
measurement, i.e. any measurement that may be achieved using passive and
active linear optics, homodyne detection and auxiliary modes prepared in
Gaussian states \cite{gez02,jal07}.
We start our discussion by reviewing the main ideas at
the basis of the definition of the discord.
Let us consider two classical random variables $A$ and $B$ with joint
probability $p_{\sab}(a,b)$; the total correlations between the two
variables are measured by the mutual information.
The latter may defined by two equivalent expressions $I(A;B)=H(A)
+ H(B) - H(A,B)$ and $I(A;B) = H(A) - H(A|B) \equiv H(B) - H(B|A)$
where $H(X)= - \sum_x p_\smx (x)\, \log p_\smx (x)$ is the Shannon entropy
of the corresponding probability distribution and the
conditional entropy is defined in terms of the conditional probability
$p_\scab (a|b)$ as $H(A|B)=-\sum_{ab} p_\sab (a,b)\,
\log p_\scab (a|b)$. The idea of quantum discord
grows out of the fact that the quantum version of the mutual
information of a bipartite state $\varrho_\sab$ may be defined in two
nonequivalent ways. The first is obtained by the straightforward
quantization of $I(A;B)$, i.e. $I(\varrho_\sab)=S(\varrho_\sma)
+ S(\varrho_\smb) - S(\varrho_\sab)$ where $S(\varrho)= -
\hbox{Tr}[\varrho\,\log\varrho]$ is the Von-Neumann entropy of
the state $\varrho$ and
$\varrho_{\sma(\smb)}=\hbox{Tr}_{\smb(\sma)}[\varrho_\sab]$,
are the partial traces over the two subsystems.
On the other hand, the quantization of the expression based
on conditional entropy, i.e. the extractable information,
involves the conditional state of a
subsystem after a measurement performed on the other one
and this fact has three relevant consequences: i) the symmetry
between the two subsystems is broken; ii) this quantity
depends on the choice of the measurement; iii) the resulting
expression is generally different from $I(\varrho_\sab)$.
Let us denote by $\varrho_{\sma k} = 1/p_\smb (k)\hbox{Tr}_\smb
[\varrho_\sab\, {\mathbb I}\otimes \Pi_k]$ with $p_\smb (k)=\hbox{Tr}_\sab
[\varrho_\sab\, {\mathbb I}\otimes \Pi_k]$, the conditional state of the
system $A$ after having observed the outcome $k$ from a measurement
performed on the system $B$. In turn, $\{\Pi_k\}$,
$\sum_k \Pi_k = {\mathbb I}$ denotes a probability operator-valued
measure (POVM) describing a generalized measurement.
The quantum analogue of the mutual information defined via the
conditional entropy is defined as the upper bound
$J_A = \sup_{\{\Pi_k\}} S(\varrho_\sma) - \sum_k p_\smb (k)
S(\varrho_{\sma k})$
taken over all the possible measurement.
Finally, the quantum A-discord is defined in terms of the mismatch $
C(\varrho_\sab) = I(\varrho_\sab) - J_\sma(\varrho_\sab)$. Analogously
one is led to define the B-discord through the entropy of conditional
states of system $B$. In the following we show that the extractable
information $J(\varrho_\sab)$ for two modes gaussian states can be
maximized over the class of Gaussian measurements, and that the mismatch
between $ C(\varrho_\sab)$  is actually minimized by a POVM rather than
a projective measurement.  As we will see it is enough to focus on
A-discord since the results for the B-discord are recovered by a
repameterization of the state: from now on we refer to A-discord as the
discord of the quantum state $\varrho$ and omit the indication of the
subsystem. Recently, a different quantity has been introduced
\cite{Wu09}, which is essentially a symmetrized version of the discord.
\par
%%%
We start our analysis by proving a general result:
quantum discord is invariant under local unitary operations,
i.e. $C(U_\sma \otimes U_\smb\,
\varrho_\smab\, U_\sma^\dag \otimes U_\smb^\dag )=C(\varrho_\smab)$,
$\forall \varrho$ and any choice of the local unitaries. The proof
simply follows by first noticing that the mutual information
$I(\varrho_\smab)$ is written in terms of two- single-system entropies
and thus it is not changed by the action of local unitaries.
Furthermore, extractable information rewrites as
$J(\varrho)=S(\varrho_\sma)-\sum_k p^\prime_\smb (k)
S(\varrho^\prime_{\sma k})$ where the primed quantities are evaluated
using the transformed POVM $\Pi^\prime_k=U_\smb^\dag \Pi_k U_\smb$.
Since this amounts to a repametrization of POVMs, which does not change
the superior, invariance is proved.
This result is relevant since it allows us to focus our
analysis on Gaussian states whose covariance matrix is in a standard form.
Indeed, let us now consider bipartite Gaussian states i.e.,
states that can be characterized by their covariance matrix
${\boldsymbol \sigma}=\left(\begin{array}{cc}A & C \\ C &
B\end{array}\right)$. By means of local unitaries
that preserve the Gaussian character of the state, i.e. local symplectic
operations, ${\boldsymbol \sigma}$ may be brought to the so-called standard form, i.e.
$A=\hbox{diag}(a,a)$, $B=\hbox{diag}(b,b)$, $C=\hbox{diag}(c_1,c_2)$.
The quantities  $I_1=\det A$, $I_2=\det B$, $I_3=\det C$,
$I_4=\det \boldsymbol{\sigma}$ are left unchanged by the transformations, and are
thus referred to as symplectic invariants.
The local invariance of the discord has therefore two main consequences.
On the one hand, $C(\varrho)$ may be written in terms of symplectic invariants only.
On the other hand, it allows us to restrict to states with
${\boldsymbol \sigma}$ already in the standard form.
In particular, while the derivation we give for the
Gaussian discord is applicable to the general case, for
the explicit calculations we will focus on the relevant subclass
of states for which $c1=-c2$, i.e the squeezed-thermal states (STS) $\varrho=S(r)\, \nu_1\otimes\nu_2\, S^\dag(r)$, where $S(r)=e^{r (a^\dag b^\dag - ab)}$
is the two-mode squeezing operator and $\nu_j=\sum_k N_j^k(1+N_k)^{-k-1}
|k\rangle\langle k|$, $j=1,2$ are chaotic states with $N_j$ average number
of thermal photons. Using this parametrization we have $a=(N_r+\frac12)+
N_1(1+N_r)+ N_2N_r$, $b=(N_r+\frac12) + N_2(1+N_r)+ N_1N_r$,
and $c_1=-c_2=(1+N_1+N_2)\sqrt{N_r(1 + N_r)}$ where $N_r=\sinh^2 r$.
\par
The definition of the Gaussian quantum discord is based on the minimization of the
mismatch $I(\varrho)-J(\varrho)$ over single mode generalized Gaussian measurements.
A first class of such POVMs may be written as \cite{gez02,jal07}
$\Pi_\smbx= D(\X) \varrho_\smm D^\dag (\X)$,
$\int\!d\X\,\Pi_\smbx = \openone$, where $\X$ is a two
dimensional real vector and $\varrho_\smm$ a generic zero mean
Gaussian state whose covariance matrix is
$\sigma_\smm=\left(\begin{array}{cc}\alpha & \gamma \\ \gamma &
\beta\end{array}\right)$, with fixed parameters
$\alpha,\beta \in {\mathbbm R}^+$, $\gamma\in
{\mathbbm R}$. If one performs
the measurement described by $\Pi_\smbx$ on, say, mode $B$ of a bipartite
Gaussian state, then the distribution of the outcomes
$p(\X)$ is a bimodal Gaussian with covariance matrix $(B+\sigma_\smm)$,
whereas the conditional state $\varrho_\smbx$ of mode $A$ is
a Gaussian state of mean $X^\smt(B+\sigma_\smm)^{-1} C^\smt$ and
covariance matrix given by the Schur complement
$\sigma_{P}= A- C(B+\sigma_\smm)^{-1} C^\smt$ \cite{jen03,tak08}.
\par
Quantum discord may be written as $C(\varrho) =
S(\varrho_\smb)-S(\varrho)+\inf_{\{\Pi_\smbx\}} \int\! d\X\, p(\X)
\,S(\varrho_\smbx)$ and the {\em general form of Gaussian quantum discord} is:
\be
C(\varrho)= h(\sqrt{I_2})-h(d_-)-h(d_+)
+\inf_{{\textstyle{\sigma_{\scriptstyle M}}}} h(\sqrt{\sigma_{P}})
\ee
where, $h[x]=(x+\frac12)\log(x+\frac12)-(x-\frac12)\log(x-\frac12)$ and $d_\pm$
are the symplectic eigenvalues of $\varrho$, expressed by
$d_\pm^2 = \frac12 \left[\Delta \pm \sqrt{\Delta^2-4 I_4}\right]$,
$\Delta=I_1+I_2+2I_3$. In deriving the expression for $C(\varrho)$,
we have used two facts: i) the
entropy of a Gaussian state depends only on the covariance matrix
and ii) the covariance matrix $\sigma_{P}$
of the conditional state does not depend on the outcome of the measurement itself.
This facts allows for a simplification of the minimization
required to obtain the final expression general expression of the Gaussian discord.  Indeed,
for the relevant case of STS, and for any choice of $N_1$, $N_2$ and $N_r$, the minimum of the mismatch $I(\varrho)-J(\varrho)$ is obtained for $\alpha=\beta=1/2$, $\gamma=0$
i.e. when the covariance matrix of the measurement is the identity. This
corresponds to the coherent state POVM, i.e. to the joint measurement of
canonical operators, say position and momentum, which may realized on
the radiation field by means of
heterodyne detection \cite{YS80}. It turns out that the same result is obtained
even if we generalize the class of Gaussian measurements to include non
covariant ones $\Pi_\smbz= D(\X) \varrho_{\smm}(\Y) D^\dag (\X)$, where now
the vector $\Z=(\X,\Y)$ includes the no longer fixed
parameters of the covariance matrix $\sigma_M=\sigma_M(\Y)$.
Indeed, since the integrand in
$\inf_{\{\Pi_\smbz\}} \int\! d\Z\, p(\Z) \,S(\varrho_\smbz)$
is always positive we have
$\inf_{\{\Pi_\smbz\}} \int\! d\Z\, p(\Z)\,S(\varrho_\smbz)\geq
\inf_{\{\Pi_\smbz\}} S(\varrho_\smbz)=\inf_{\Y} h(\sqrt{\sigma_{P}(\Y)})$ and the above results apply for any $\Z$.  Upon substituting
$\sigma_\smm \rightarrow {\mathbbm I}/2$,
we can now explicitly write the Gaussian discord for the generic bipartite STS in terms of
of symplectic invariants as:
\be
C(\varrho) = h(\sqrt{I_2})-h(d_-)-
h(d_+)+h(\frac{\sqrt{I_1}+2
\sqrt{I_1 I_2} + 2 I_3}{1+2 \sqrt{I_2}}).
\ee
Upon exchanging $I_1 \leftrightarrow I_2$ one can pass from the
$A$-discord to the $B$-discord.
\\We are now ready to start our discussion about the properties and the
operational meaning of Gaussian quantum discord. At first we notice that
$C(\varrho)\neq0$ as far as $N_r\neq0$. Given that Gaussian states in
standard form are separable for $N_r \leq N_1 N_2/(1+N_1+N_2)$, this
confirms that for CV Gaussian states there are separable states with
nonzero discord. Besides, since $N_r \neq 0 \Leftrightarrow c \neq 0$ we
have that bipartite Gaussian states have always nonzero Gaussian
discord, except when they are product states. The same condition
characterizes the class of tomographically faithful states for
reconstruction of quantum operations \cite{max05}, and this provides an
operational meaning for the quantum correlations in separable states with
positive discord.  The behavior of $C(\varrho)$ for small and large
$N_r$ is given by
$C(\varrho)\stackrel{N_r\ll 1}{\simeq}f_1(N_1,N_2) N_r$ and
$C(\varrho)\stackrel{N_r\gg 1}{\simeq}f_2(N_1,N_2) + f_3 (N_1,N_2) \log N_r$
respectively, where $f_1$ is a decreasing function of $N_2$ at any fixed
value of $N_1$ and $f_2$, $f_3$ are decreasing functions of both $N_1$
and $N_2$.
%%%%%%%%%%%%%
\begin{figure}[h!]
\includegraphics[width=0.38\columnwidth]{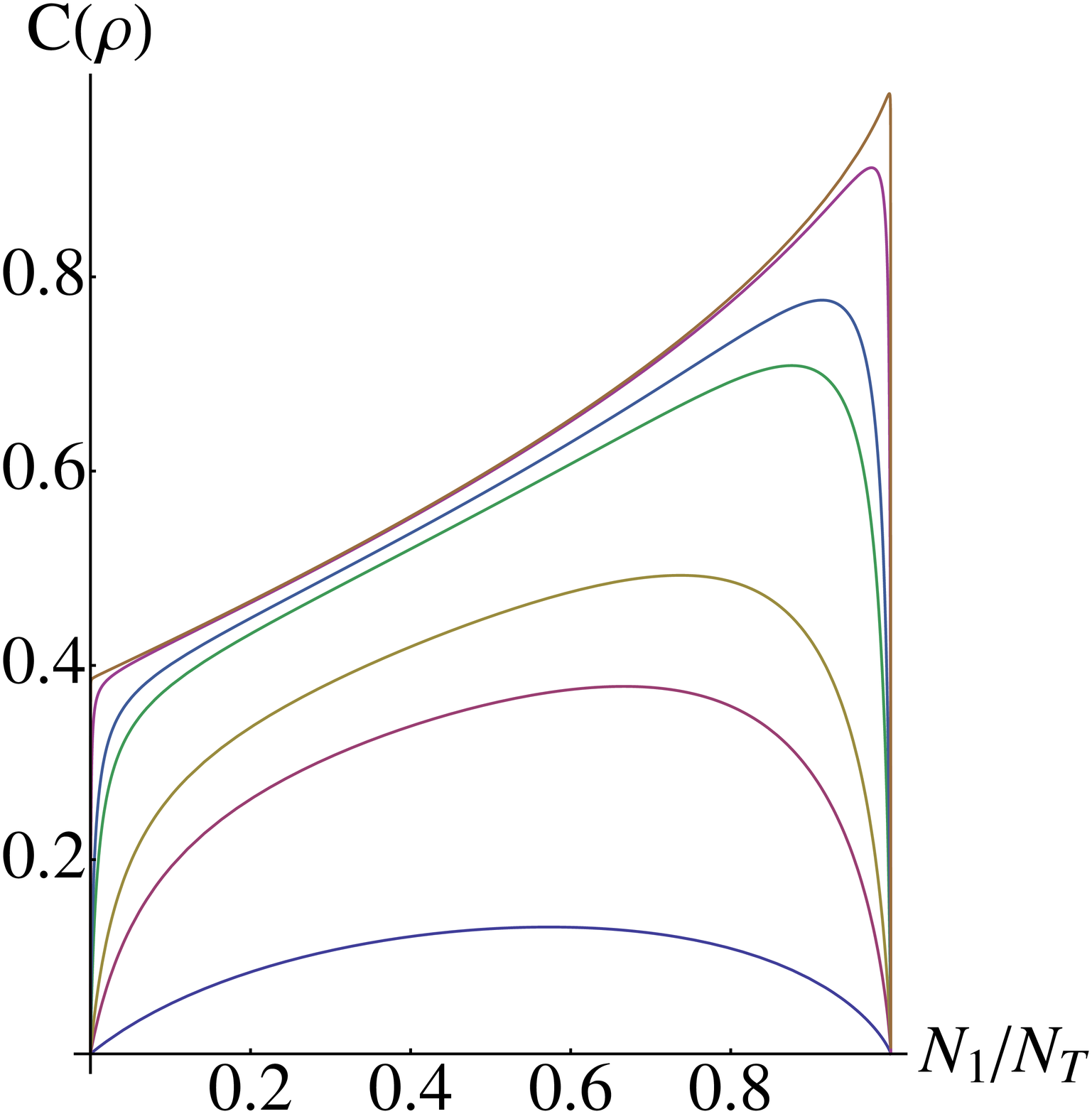}
\includegraphics[width=0.38\columnwidth]{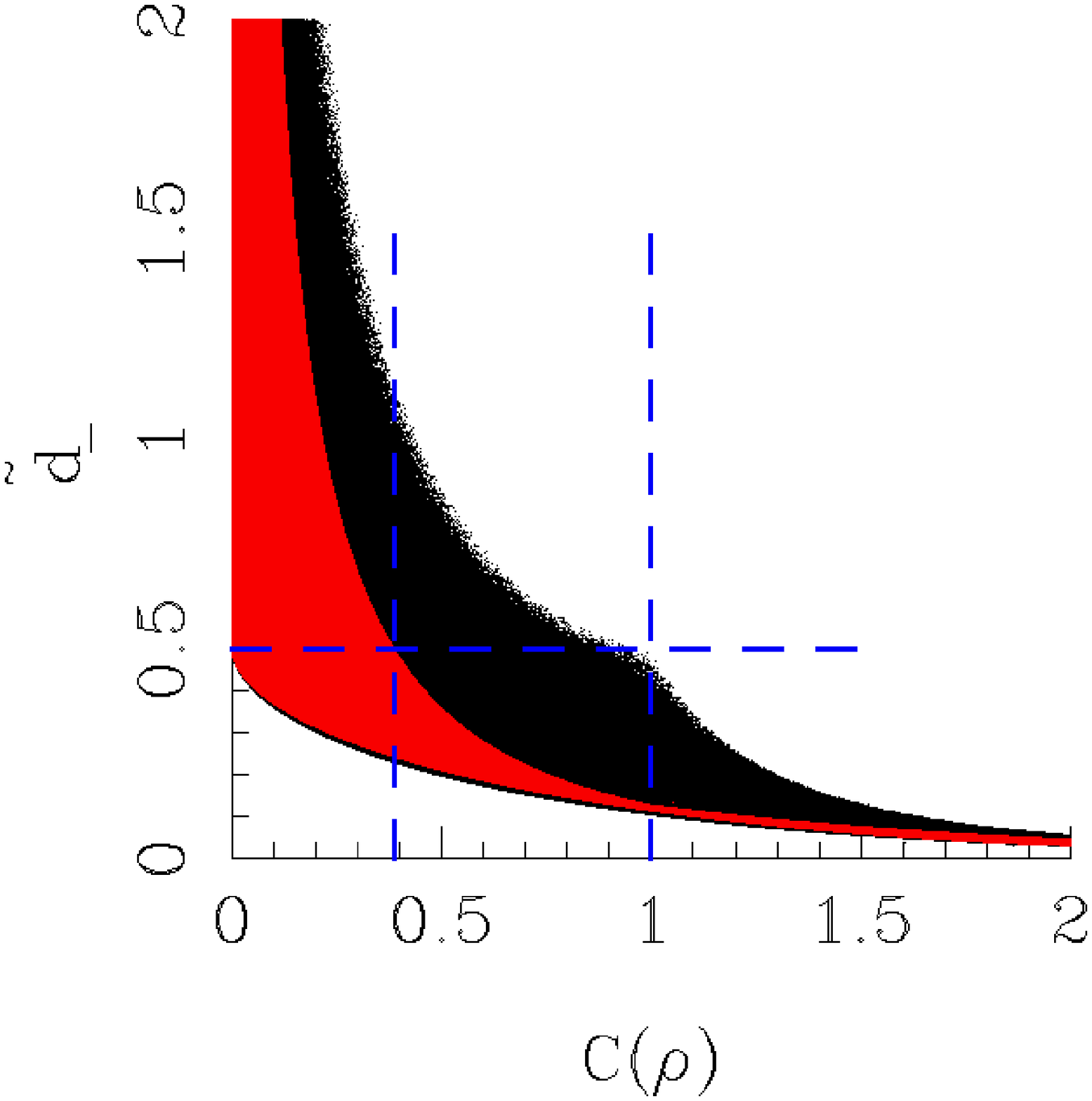}
\caption{(color online) Left: Gaussian discord $C(\varrho)$ for STS at separability
threshold as a function of the ratio $N_1/N_\smt$ for increasing  values
of the total energy $N_\smt$ of the states
(from bottom to top $N_\smt=1, 5, 10, 50, 10^2, 10^3, 10^5$).
Right: symplectic eigenvalues of the partial transpose $\tilde{d}_-$
versus $C(\varrho)$ for randomly generated STS.
The red region corresponds to symmetric states.
We also report the separability threshold $\tilde{d}_-=\frac12$ and the
corresponding threshold $C(\varrho)=1$ ($C(\varrho)=2 \log 2-1$ symmetric STS).
\label{f:GQD}}
\end{figure}
\par
%%%%%%%%%%%%%
We now focus our attention on how $C(\varrho)$ relates with other meaningful properties of the states. In Fig. \ref{f:GQD}a we report $C(\varrho)$ at the separability
threshold $N_r = N_1 N_2/(1+N_1+N_2)$, as a function of the ratio
$N_1/N_\smt$ for increasing (from bottom to top) values of
$N_\smt=a+b-1$, which is the total energy of the Gaussian state under
investigation ($N_\smt=N_1+N_2+2N_1 N_2$ at separability threshold). The plot
suggests two important facts. First, Gaussian discord is an increasing function
of the total energy, and is maximized when most of the thermal photons
are placed on the unmeasured system, thus maximizing the purity of
the measured one. Second, the {\it Gaussian discord for separable states
is always smaller than one}. The existence of a bound has
been confirmed numerically by the random generation of a large
number of bipartite Gaussian states in the standard form: in Fig.
\ref{f:GQD}b, we report the smaller symplectic eigenvalue $\tilde{d}_-$
of the partially transposed state, obtained by replacing $I_3\rightarrow
-I_3$ in the the formula for $d_-$, as a function of Gaussian
discord. Since a Gaussian state is entangled iff $\tilde{d_-}<\frac12$
we have that for $0\leq C(\varrho)\leq 1$ we have either separable or
entangled states, whereas all the states with $C(\varrho)>1$ are
entangled.
\par
\begin{figure}[h!]
\includegraphics[width=0.4\columnwidth]{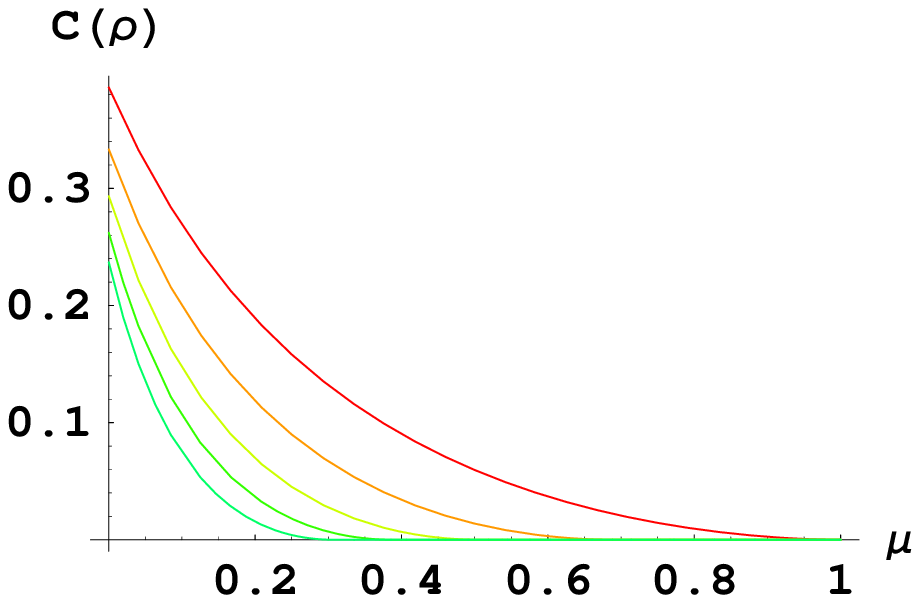}
\includegraphics[width=0.4\columnwidth]{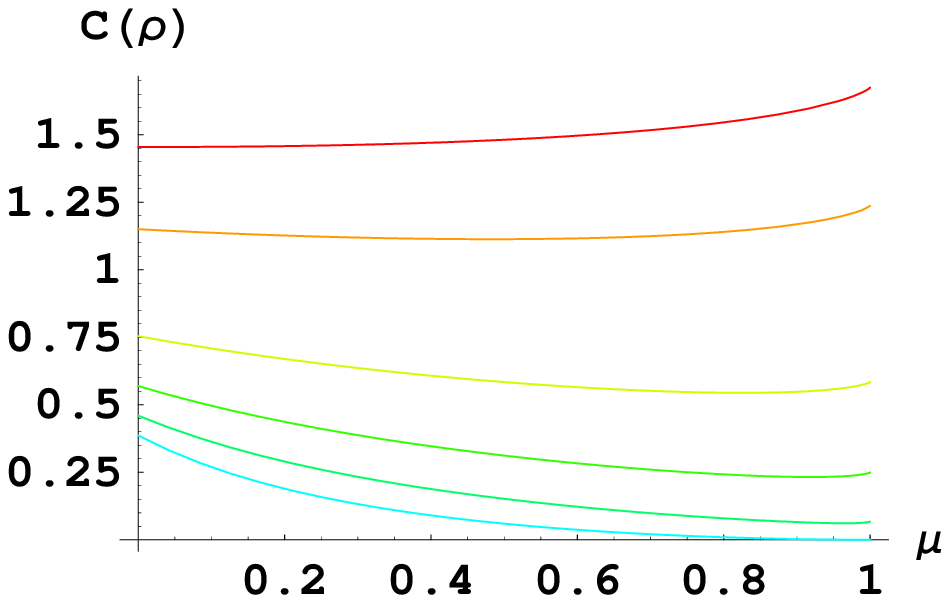}
\caption{(color online) Gaussian discord $C(\varrho)$ for symmetric
STS ($N_1=N_2=N_s$) as a function of the global purity $\mu$ of the state
 $\rho$ and the smaller symplectic eigenvalue
of its partially transpose $\tilde{d}_-$.  Left:
separable states; from bottom to top
$\tilde{d}_-=0.5,0.6,0.7,0.8,0.9$; $C(\varrho)$ monotonically
decreases with $\mu$. Right: entangled states ; from bottom to top
$\tilde{d}_-=0.5,0.4,0.3,0.3,0.1,0.06284$; $C(\varrho)$ monotonically
increases with $\mu$ only for $\tilde{d}_-\ge 0.06284$.
\label{Fig:CVDmudent}}
\end{figure}

The relation between the discord and the entanglement can be further clarified
by analyzing the case of symmetric STS, i.e., $N_1=N_2=N_s$.
Here we focus on the behavior of $\,C(\varrho)$ with respect to global
purity of the state $\mu=(1+2 N_s)^{-2}$ and $\tilde{d}_-=e^{-2 s}(1+2
N_s)/2$. A first important observation is that for {\it fixed purity
$\,C(\varrho)$ turns out to be a growing function of the entanglement},
whereas at fixed values of
$\tilde{d}_-$ the behavior is more involved. In Fig. \ref{Fig:CVDmudent}
we plot $\,C(\varrho)(\mu,\tilde{d}_-)$ at fixed values of  $\tilde{d}_-$. We
can distinguish two different cases. For non entangled states
($\tilde{d}_-\ge1/2$), $\,C(\varrho)$ decreases with $\mu$, and it thus
is an increasing function of the total energy of the state
$N_T=\tilde{d}_- -1+(4 \tilde{d}_- \mu)^{-1}$. The limiting value is
thus reached at infinite energy and the latter is in general given by $C(\varrho)(\mu \rightarrow 0,\tilde{d}_-)=(1+2
\tilde{d}_-)\ln{[(1+2 \tilde{d}_-)/ \tilde{d}_-)]}-(1+(1+2 \tilde{d}_-)\ln{2})$. Therefore, for non entangled symmetric states
$C(\varrho)\le 2 \log 2-1$; the latter bound is also reported in Fig.
\ref{f:GQD}, and defines the limit of the red region corresponding to
symmetric separable states with non zero discord.
%%%%%%%%%%%%
%%%%%%%%%%%%
\\As for the entangled states ($\tilde{d}_-\le1/2$), the behavior of
$C(\varrho)$ with $\mu$ is more complex. For states which are highly
entangled ($\tilde{d}_-\le 0.06284$) the discord decreases (grows)
monotonically with $\mu\,\, (N_T)$. Indeed, $J(\varrho)\approx
h(\sqrt{I_1})$ when $\tilde{d}_-\rightarrow 0$, i.e, the extractable
information is maximized, and $C(\varrho)$ is maximum for pure states.
For intermediate values of the entanglement,
$\tilde{d}_-\in (0.06284,0.5)$, $C(\varrho)$ has a non monotonic
behavior with $\mu \,\, (N_T)$. In particular, the maximum discord
is reached for $\mu=1$ (pure states) only for $\tilde{d}_- \geq 0.1282$,
while its minimum is reached for intermediate values of $\mu$ that depend the
actual value of $\tilde{d}_-$.
The overall non monotonic behavior of $C(\varrho)$ corresponds
to a situation in which, at fixed value of entanglement, the quantumness
of the state as measured by the Gaussian discord, varies depending on
the total correlations present in the state, and consequently the
ordering of the states with respect to their quantumness significantly
differs by that given by the entanglement.  We also emphasize that by
fixing the value of $\tilde{d}_-$ one also fixes the value of the
teleportation fidelity $F=(1+2\tilde{d}_-)^{-1}$ of coherent states
\cite{fur98}. This means that by varying the global purity of the state $\rho$
shared by Alice and Bob, the same fidelity can be achieved with
different quantum resources as measured by the Gaussian discord.
\par
We finally address the fundamental issue of the evolution of quantum discord
in noisy channels. Let us consider bipartite Gaussian states that evolve
according to Lindblad Master equation $\dot\varrho = \frac12
\sum_j \Gamma_j M_j L[a]\varrho + \Gamma_j (1+M_j) L[a^\dag]\varrho$,
which describes the Markovian interaction of the two modes with independent
thermal reservoirs, $\Gamma_j$ and $M_j$ being the damping factor and
the average number of  thermal photons of the two reservoirs respectively.
The mapping induced by the ME is Gaussian and the covariance matrix of
the evolved state is $\boldsymbol{\sigma}_t =
\boldsymbol{\Gamma}^\frac12_t
\boldsymbol{\sigma}\boldsymbol{\Gamma}^\frac12_t + (1-
\boldsymbol{\Gamma}_t
)\boldsymbol{\sigma}_\infty$, where $\boldsymbol{\Gamma}_t = \bigoplus_j
e^{-\Gamma_j t} \id_2$ and
$\boldsymbol{\sigma}_\infty=\hbox{Diag}(M_1+\frac12, M_1+\frac12,
M_2+\frac12, M_2+\frac12)$ is the covariance matrix of the reservoir,
which also describes the stationary state of the system.
If $\boldsymbol{\sigma}_{t=0} $ is in standard form
its parameters evolve as:
$a^\prime= a\, e^{-\Gamma_1 t} + (1-e^{-\Gamma_1 t}) (M_1 +\frac12)$,
$b^\prime= b\, e^{-\Gamma_2 t} + (1-e^{-\Gamma_2 t}) (M_2 +\frac12)$,
$c^\prime= c\, e^{-\frac12 (\Gamma_1 + \Gamma_2)t}$, i.e.
$a^\prime>a$, $b^\prime>b$, and  $c^\prime<c$. Since $C(\varrho)$ is
a decreasing function of $a$ and $b$ and an increasing function of $c$
we have that Gaussian discord monotonically decreases in noisy channels.
On the other hand, it has been shown that the decrease should be smooth
since an arbitrary Markovian evolution can never lead to a sudden
disappearance of discord \cite{Ale09}.
An open question remains the effect
of non-Markovian dynamics, which has been proved to
produce oscillations in the dynamics of Gaussian entanglement
\cite{sbm07,ruv09}. We also expect Gaussian discord to
increase if the two parties interact with a common reservoir
\cite{ben03}.
%{\bf An open question remains the effect of non-Markovian dynamics \cite{sbm07,ruv09}
%and we also expect Gaussian discord to increase if the two parties interact with a common %reservoir \cite{ben03}}.
\par
In conclusion, in this Letter we have extended the notion of the discord \cite{OZ01}  to
continuous variable systems and discuss its properties. 
We have defined the Gaussian discord $C(\varrho)$ for two modes Gaussian states
and we have shown the general analytical procedure to derive it. In particular,
for the relevant subclass of squeezed thermal states
(STS), we have shown that the extractable information is maximized by a generalize measurement
i.e. the coherent state POVM corresponding to heterodyne detection. Just as the entanglement, $C(\varrho)$ is invariant under local
unitary operations and it is zero only for (thermal) product states.
For separable states $C(\varrho)$ grows with the total energy and it is bounded.
Numerical evidences show that in general $C(\varrho)<1$ while analytical
calculations show that for separable symmetric STS the bound reduces to $C(\varrho)<2
\ln{2}-1$. For symmetric STS we have also shown that the behavior of
$C(\varrho)$  strongly depends on the amount of entanglement present in
the state: it increases with the total purity only when the entanglement
is large, whereas it shows a richer behavior for smaller values of
entanglement. Our results pave the way for the general discussion about the 
quantum discord in continuous variable systems and for its the experimental determination with current technology.
%%%%%%%%%%%%%%%%%%%%%%%%%%%%%%%%%%

%%%%%%%%%%%%%%%%%%%%%%%%%%%%%%%%%%%
\end{document}